\documentstyle[12pt,aasms4]{article}

\begin{document}

\title{Measuring the galaxy power spectrum with multiresolution
decomposition -- IV. redshift distortion}

\author{XiaoHu Yang\altaffilmark{1,2}\quad Long-Long
Feng\altaffilmark{1,2,3}\quad YaoQuan Chu\altaffilmark{1,2}\quad
Li-Zhi Fang\altaffilmark{4}}

\altaffiltext{1}{Center for Astrophysics, University of Science
and Technology of China, Hefei, Anhui 230026,P.R.China}
\altaffiltext{2}{National Astronomical Observatories, Chinese
Academy of Science, Chao-Yang District, Beijing, 100012, P.R.
China} \altaffiltext{3}{Institute for Theoretical Physics,
Academic of Science, Beijing, 100080, P.R. China
}\altaffiltext{4}{Department of Physics, University of Arizona,
Tucson, AZ 85721}

\begin{abstract}

In this paper, we develop a theory of redshift distortion of the
galaxy power spectrum in the discrete wavelet transform (DWT)
representation. Because the DWT power spectrum is dependent of
both the scale and shape (configuration) of the decomposition
modes, it is sensitive to distortion of shape of the field. On the
other hand, the redshift distortion causes a shape distortion of
distributions in real space with respect to redshift space.
Therefore, the shape-dependent DWT power spectrum is useful to
detect the effect of redshift distortion. We first established the
mapping between the DWT power spectra in redshift and real space.
The mapping depends on the redshift distortion effects of (1) bulk
velocity, (2) selection function and (3) pairwise peculiar
velocity. We then proposed $\beta$-estimators using the DWT
off-diagonal power spectra. These $\beta$-estimators are model-free 
even when the non-linear redshift distortion effect is not negligible.
Moreover, these estimators do not rely
on the assumption of whether the pairwise velocity dispersion
being scale-dependent. The tests with N-body simulation samples
show that the proposed $\beta$-estimators can yield reliable
measurements of $\beta$ with about 20\% uncertainty for all
popular dark matter models. We also develop an algorithm for
reconstruction of the power spectrum in real space from the
redshift distorted power spectrum. The numerical test also shows
that the real power spectrum can be well recovered from the
redshift distorted power spectrum.

\end{abstract}


\section{Introduction}

In three previous papers we have developed a method of measuring
galaxy power spectrum with discrete wavelet transform (DWT)
decomposition, which is an alternative of the Fourier power
spectrum detection (Fang \& Feng, 2000 (paper I), Yang et al 2001a
(paper II), 2001b (paper III)). The DWT power spectrum estimator
is information-loseless, and optimized in the sense that the
spatial resolution is adaptive automatically to the scale to be
studied. A test with observed sample of the LCRS galaxies showed
that the DWT estimator can give a robust measurement of the power
spectrum.

In this paper, we continue our effort in this direction. The topic
this time is to develop theory and algorithm of redshift
distortion in the DWT representation.

In terms of power spectrum measurement, the central problem of
redshift distortion is to find the mapping between the power
spectra in redshift and physical spaces. In our paper II, we have
already studied this mapping. However, that mapping is directly
obtained by a wavelet transform of the mapping of the power
spectrum in the Fourier representation (Peacock \& Dodds 1994).
Although a Fourier mode can be transformed into DWT modes and {\it
vice versa}, we should be careful in doing the transform of a
second order statistical quantity from the Fourier representation
into the DWT one. As we have showed in paper I, the covariance of
density contrast in the DWT representation $\langle \epsilon_{\bf
j,l}\epsilon_{\bf j',l'} \rangle$ is not equivalent to the Fourier
counterpart $\langle \hat{\delta}_{\bf k}\hat{\delta}_{\bf k}
\rangle$, because the Fourier mode is subjected to the central
limit theorem, while mode localized in both scale and physical
space may not be so. This is because the DWT mode is characterized
by not only the scale, but also the configuration (or shape) of
the mode. Simply speaking, the variance $\langle \epsilon_{\bf
j,l}\epsilon_{\bf j',l'} \rangle$ contains information of both the
scale and phase (shape), while $\langle \hat{\delta}_{\bf
k}\hat{\delta}_{\bf k} \rangle$ does not contain information of
phases (Fang \& Thews 1998).

We have found in paper II that, the Fourier mapping of redshift
distortion can be employed for diagonal DWT power spectrum, but
not off-diagonal DWT power spectrum. This is due to the
shape(phase)-dependence of DWT modes. A 3-D Fourier mode with
wave-vector ${\bf k}=(k_1,k_2,k_3)$ can be transformed into mode
${\bf k'}=(k'_1,k'_2,k'_3)$ by a coordinate rotation as long as
$k^2_1+k^2_2+k^2_3 =k'^2_1+k'^2_2+k'^2_3$. Therefore, for an
isotropic random field, the Fourier modes with the same $k=|{\bf
k}|$ are statistically equivalent. However, the DWT modes do not
share the same property. The length scale of a 3-D wavelet mode
with scale index ${\bf j}=(j_1,j_2,j_3)$ is
$L/[2^{2j_1}+2^{2j_2}+2^{2j_3}]^{1/2}$, where $L$ is the length
scale of the sample. Generally, one cannot transform a mode $(j_1,
j_2, j_3)$ to $(j'_1, j'_2, j'_3)$ by a rotation, even if they
have the same scale.

As a consequence, the redshift distortion in the DWT presentation
will be shape-dependent. For instance, in the Fourier
representation, the linear redshift distortion factor $(1+\beta
\mu^2)^2$ (Kaiser 1987), where $\beta$ is the so-called redshift
distortion parameter, depends only on $\mu=k_3/k$, i.e. the
cosine of the angle between the wavevector $\bf k$ and the
line-of-sight. While in the DWT representation, the linear
redshift distortion factor will depend on not only the scale, but
also the shape of DWT modes.

An important application of redshift distortion is to determine
the redshift distortion parameter $\beta$ (e.g. Hamilton 1998), which 
contains valuable information of the cosmological
mass density parameter and the bias of galaxies. The
shape-dependence of redshift distortion is very useful for the
parameter determination. We will develop an algorithm of $\beta$
estimation with diagonal and off-diagonal DWT power spectra.

Moreover, the DWT representation provides an easy way to study the
effect of selection function on the redshift distortion. In the
first three papers, we assumed that the selection function
$\bar{n}({\bf x})$ is known. Actually, the ``known" is not
necessary. According to the definition, selection function is an
observed galaxy distribution if galaxy clustering is absent.
Therefore, for a consistent algorithm of power spectrum, the
selection function should be and can be obtained by the galaxy
distribution itself. We will show that in the DWT representation
the selection function can be determined by the distribution of
galaxies without other assumption. This algorithm  is convenient
to estimate the contribution of selection function to redshift
distortion.

The paper will be organized as follows. \S 2 is a summary of the
algorithm of the DWT power spectrum. In \S 3, we develop a theory
of the redshift distortion with a multiresolution analysis. The
emphases are the real-redshift mapping of diagonal and
off-diagonal power spectrum. We proposed $\beta$-estimators,
which are tested by N-body simulation samples of popular dark
matter models (\S 4 and 5). Finally, the conclusions and
discussions will be presented in \S 6. The mathematical stuffs
with the relevant DWT quantities are given in Appendix.

\section{Algorithm of the DWT power spectrum}

\subsection{The DWT power spectrum}

We summarize the algorithm of the DWT power spectrum. The details
can be found  in Fang \& Feng (2000) and Yang et al (2001).

If the position measurement is perfectly precise, the number
density distribution of galaxies can be written as
\begin{equation}
n^g({\bf x})=\sum_{i=1}^{N_g}w_i\delta_D({\bf x-x}_i)=
\bar{n}({\bf x})[1+\delta({\bf x})]
\end{equation}
where $N_g$ is the total number of galaxies, ${\bf x_i}$ is
position of $i$th galaxy, $w_i$ is its weight, and $\delta_D$ the
3-D Dirac $\delta$ function. $\bar{n}({\bf x})$ is selection
function given by the mean number density of galaxies when galaxy
clustering is absent, and $\delta({\bf x})$ is the density
contrast fluctuation in the underlying matter distribution.

Without loss of generality, we enclose the sample in a cubic box
with side lengths $(L_1,L_2,L_3)$. In the DWT representation,
$n^g({\bf x})$ is decomposed into
\begin{equation}
n^g({\bf x}) = n^{({\bf j})}({\bf x}) +
\sum_{n=1}^{7}\sum_{j'_1=j_1}^{\infty}\sum_{j'_2=j_2}^{\infty}
\sum_{j'_3=j_3}^{\infty}\sum_{l_1=0}^{2^{j'_1}-1}
\sum_{l_2=0}^{2^{j'_2}-1}\sum_{l_3=0}^{2^{j'_3}-1}
\tilde{\epsilon}^{g,n}_{\bf j',l} \psi_{\bf j',l}^{(n)}({\bf x}),
\end{equation}
where, $n^{({\bf j})}({\bf x})$ is given by
\begin{equation}
n^{({\bf j})}({\bf x})=
\sum_{l_1=0}^{2^{j_1}-1}\sum_{l_2=0}^{2^{j_2}-1}\sum_{l_3=0}^{2^{j_3}-1}
 \epsilon^{g}_{\bf j,l}\phi_{\bf j,l}({\bf x}).
\end{equation}
The index ${\bf j}=(j_1,j_2,j_3)$ stand for 3-D scales
$L_2/2^{j_1},L_2/2^{j_2},L_3/2^{j_3}$, and index ${\bf
l}=(l_1,l_2,l_3)$ for the position of cell $l_1L_1/2^{j_1} < x_1
\leq (l_1+1)L_1/2^{j_1}$, $l_2L_2/2^{j_2} < x_2 \leq
(l_2+1)L_2/2^{j_2}$, $l_3L_3/2^{j_3} < x_3 \leq
(l_3+1)L_3/2^{j_3}$.

In eq.(3), $\epsilon^{g}_{\bf j,l}$ is called scaling function
coefficient (SFC) of $n^g({\bf x})$. They are given by
\begin{equation}
\epsilon^{g}_{\bf j,l} =
 \int n^g({\bf x})\phi_{\bf j,l}({\bf x})d{\bf x}=
\sum_{i=1}^{N_g} w_i \phi_{\bf j,l}({\bf x}_i),
\end{equation}
where we have used equation (1) in the last step. The scaling functions 
$\phi_{\bf j,l}({\bf x})$ play the role of window  functions for the cell
with volume $(L_1/2^{j_1})\times (L_2/2^{j_2})\times (L_3/2^{j_3})$, and 
located at position ${\bf l}$. Therefore, the SFC $\epsilon^{g}_{\bf j,l}$ 
is the mean of field $n^g({\bf x})$ in the volume 
$(L_1/2^{j_1})\times (L_2/2^{j_2})\times (L_3/2^{j_3})$ 
at position ${\bf l}$. Thus, the term $n^{({\bf j})}({\bf x})$ of eqs.(2) 
and (3) actually is the smoothed $n^{({\bf j})}({\bf x})$ by window 
on scale ${\bf j}$. 

The 3-D scaling functions $\phi_{\bf j,l}({\bf x})$ can be 
constructed by a direct product of 1-D scaling functions, i.e.
\begin{equation}
 \phi_{\bf j,l}({\bf x})=\phi_{j_1,l_1}(x_1)
  \phi_{j_2,l_2}(x_2)\phi_{j_3,l_3}(x_3).
\end{equation}

In eq.(2), $\psi_{\bf j,l}^{(n)}({\bf x})$ are wavelets.  An 1-D wavelet $\psi_{j,l}(x)$ is the modes used to extract the fluctuations of 
a 1-D field $n(x)$ on scale $L/2^j$ at position ${\bf l}$. The 3-D wavelets 
$\psi_{\bf j,l}^{(n)}({\bf x})$ are given by mixed direct products of 
1-D scaling functions $\phi_{j,l}({x})$ and wavelets $\psi_{j,l}({x})$ 
as follows, 
\begin{eqnarray}
\psi_{\bf j',l}^{(1)}({\bf x})&=&\phi_{j'_1,l_1}(x_1)
\phi_{j'_2,l_2}(x_2)\psi_{j'_3,l_3}(x_3)\delta_{j'_1,j_1}\delta_{j'_2,j_2}
\nonumber \\
\psi_{\bf j',l}^{(2)}({\bf x})&=&\phi_{j'_1,l_1}(x_1)
\psi_{j'_2,l_2}(x_2)\phi_{j'_3,l_3}(x_3)\delta_{j'_1,j_1}\delta_{j'_3,j_3}
\nonumber \\
\psi_{\bf j',l}^{(3)}({\bf x})&=&\psi_{j'_1,l_1}(x_1)
\phi_{j'_2,l_2}(x_2)\phi_{j'_3,l_3}(x_3)\delta_{j'_2,j_2}\delta_{j'_3,j_3}
\nonumber \\
\psi_{\bf j',l}^{(4)}({\bf x})&=&\phi_{j'_1,l_1}(x_1)
   \psi_{j'_2,l_2}(x_2)\psi_{j'_3,l_3}(x_3)\delta_{j'_1,j_1} \\
\psi_{\bf j',l}^{(5)}({\bf x})&=&\psi_{j'_1,l_1}(x_1)
   \phi_{j'_2,l_2}(x_2)\psi_{j'_3,l_3}(x_3)\delta_{j'_2,j_2} \nonumber \\
\psi_{\bf j',l}^{(6)}({\bf x})&=&\psi_{j'_1,l_1}(x_1)
   \psi_{j'_2,l_2}(x_2)\phi_{j'_3,l_3}(x_3)\delta_{j'_3,j_3} \nonumber \\
\psi_{\bf j',l}^{(7)}({\bf
x})&=&\psi_{j'_1,l_1}(x_1)\psi_{j'_2,l_2}(x_2)\psi_{j'_3,l_3}(x_3).
\nonumber 
\end{eqnarray}
Accordingly, $\psi_{\bf j',l}^{(n)}({\bf x})$ with $n=1,2,3$ describe 
2-D projected fluctuations on a slice, and $n=4,5,6$ 1-D
fluctuations along a line. $\psi_{\bf j',l}^{(7)}({\bf x})$ describes 
the 3-D fluctuations inside the cell ${\bf l}=(l_1,l_2,l_3)$ on a 
scale $j'$.

Hereafter, we consider the projection of density fluctuations
onto the space spanned by $\psi_{\bf j',l}^{(7)}({\bf x})$ only.
For simplicity, we ignore the upper index $n=7$ in the notations
of $\psi_{\bf j',l}^{(n)}({\bf x})$ and
$\tilde{\epsilon}^{g,n}_{\bf j',l}$ without confusion.
Thus, the wavelet function coefficient (WFC) 
$\tilde{\epsilon}^{g}_{\bf j,l}$ of $n^g({\bf x})$ is given by
\begin{equation}
\tilde{\epsilon}^{g}_{\bf j,l}=\int n^g({\bf x})
   \psi_{\bf j,l}({\bf x})d{\bf x}=
 \sum_{i=1}^{N_g} w_i \psi_{\bf j,l}({\bf x}_i),
\end{equation}
Therefore, the WFC $\tilde{\epsilon}^{g}_{\bf j,l}$ is the amplitude
of the fluctuations of the field $n^g({\bf x})$ on scale  
$(L_1/2^{j_1})\times (L_2/2^{j_2})\times (L_3/2^{j_3})$ 
at position ${\bf l}$. 

The functions $\phi_{\bf j,l}({\bf x})$ and $\psi_{\bf
j,l}^{(n)}({\bf x})$ form a complete and orthonormal basis in
space of functions defined in 3-dimensional coordinate space. The 
decomposition of eq.(2) is information loseless. 

The DWT power spectrum of $\delta({\bf x})$ can be estimated by
\begin{equation}
P_{\bf j} = \left \langle
 \frac {[\tilde{\epsilon}^{g}_{\bf j,l}]^2}{n^2({\bf j,l})}\right \rangle
 -\left \langle \frac{1}{n({\bf j,l})}\right \rangle,
\end{equation}
where $\langle...\rangle$ is the average over ensemble. The first
term on the r.h.s. is the normalized power from $n^g({\bf x})$,
and the second term corrects for the Poisson noise. The
normalization factor $n({\bf j,l})$ is the mean selection function
in the mode $({\bf j,l})$, i.e., the number density of galaxies
in the mode $({\bf j,l})$ when galaxy clustering is absent. As has
been shown in paper I (Fang \& Feng 2000), the factor $n({\bf
j,l})$ can be absorbed into the weight factor by
\begin{equation}
\frac{n^g({\bf x})}{\bar{n}({\bf x})} =
\sum_{i=1}^{N_g}\frac{1}{\bar{n}({\bf x}_i)}w_i\delta_D({\bf
x-x}_i).
\end{equation}
Thus, equation(8) yields
\begin{equation}
P_{\bf j} = \frac{1}{2^{j_1+j_2+j_3}}\sum_{l_1=0}^{2^{j_1}-1}
\sum_{l_2=0}^{2^{j_2}-1}\sum_{l_3=0}^{2^{j_3}-1}
[\tilde\epsilon_{\bf j,l}]^2 -
\frac{1}{2^{j_1+j_2+j_3}}\sum_{l_1=0}^{2^{j_1}-1}
\sum_{l_2=0}^{2^{j_2}-1}\sum_{l_3=0}^{2^{j_3}-1}
\int\frac{\psi_{\bf j,l}^2({\bf x})}{\bar{n}({\bf x})} d{\bf x},
\end{equation}
where the WFC $\tilde\epsilon_{\bf j,l}$ is defined as
\begin{equation}
\tilde\epsilon_{\bf j,l} = \int \frac{n^g({\bf x})}{\bar{n}({\bf
x})} \psi_{\bf j,l}({\bf x})d{\bf x}
 = \sum_{i=1}^{N_g}\frac{1}{\bar{n}({\bf x}_i)} w_i \psi_{\bf
j,l}({\bf x}_i)
\end{equation}

\subsection{Selection function in the DWT representation}

By definition eq.(1), selection function $\bar{n}({\bf x})$ is an
observed galaxy distribution if galaxy clustering $\delta({\bf
x})$ is absent. That is, equation (1) requires to decompose an
observed distribution $n^g({\bf x})$ into two parts: the
``background'' $\bar{n}({\bf x})$, and the fluctuations
$\delta({\bf x})$ upon the background.

This decomposition is not easy if the selection function
$\bar{n}({\bf x})$ is position-dependent, i.e. it will mix with
the fluctuations to be detected. With the DWT analysis, one is
capable of performing this decomposition by a scale-by-scale
analysis.

When we study the fluctuations on a scale ${\bf j}$, all ${\bf
x}$-dependencies of $n^g({\bf x})$ on scales larger than this
scale play the role as a background. Thus, in terms of the scale
${\bf j}$, the background is given by a smooth of $n^g({\bf x})$
on the scale ${\bf j}$, i.e. it does not contain information of
fluctuations on scales equal to or less than ${\bf j}$.

This background has already be found from equations (2) and (3).
Because of the orthogonal relation
\begin{equation}
\int \phi_{\bf j,l}({\bf x})\psi_{\bf j',l'}({\bf x})d{\bf x}=0,
\hspace{3mm} {\rm if\ one\ of}\ j'_i\ (i=1,2,3)\ {\rm satisfies}
\ \  j'_i \geq j_i,
\end{equation}
the function $n^{({\bf j})}({\bf x})$ does not contain any
information of fluctuations of modes $({\bf j',l})$ with $j'_i
\geq j_i$. Thus, to detect the fluctuation power on the scale
${\bf j}$, the function $n^{({\bf j})}({\bf x})$ is recognized as
a background field. One can identify the selection function as
$n^{({\bf j})}({\bf x})$, i.e.
\begin{equation}
 \bar{n}^{({\bf j})}({\bf x})=
 \sum_{l_1=0}^{2^{j_1}-1}\sum_{l_2=0}^{2^{j_2}-1}\sum_{l_3=0}^{2^{j_3}-1}
    \epsilon^{g}_{\bf j,l}\phi_{\bf j,l}({\bf x}),
\end{equation}
where the superscript ${\bf j}$ means that this ``selection
function'' is only for the scale ${\bf j}$.

In the plane parallel approximation, selection function depends on
$x_3$ only, i.e. the coordinate in redshift direction. From
equation (13), the selection function is given by
\begin{equation}
\bar{n}(x_3)= n^{(00j_3)}({x_3}) = \sum_{l_3=0}^{2^{j_3}-1}
 \epsilon^{g}_{00j, l}\phi_{j_3,l}(x_3).
\end{equation}
 From equation (6), the SFC $\epsilon^{g}_{00j, l}$ is actually
given by an average of $n^g({\bf x})$ over the plane $(x_1,
x_2)$, and decomposition along $x_3$ direction.

Using equation (14), the WFC $\tilde\epsilon_{\bf j,l}$ is now
calculated by
\begin{equation}
 \tilde\epsilon_{\bf j,l}
  = \sum_{i=1}^{N_g}\frac{1}{n^{(00j_3)}({x_3}_i)} w_i
    \psi_{\bf j,l}({\bf x}_i).
\end{equation}
which presents a simple algorithm for deriving the selection
function from observed galaxy samples.

To test this algorithm, we produce mock galaxy samples using
N-body simulation, of which the details will be given in \S 4.
For simplicity, the selection effect is applied along one axis
(e.g., $x_3$ direction) of 3-dimensional Cartesian coordinates
under the plane parallel approximation. For the simulation sample
in the cubic box with a side length of 256 h$^{-1}$ Mpc, we
replicate the sample along $x_3$ direction, and choose the mock
galaxies located between 100-356 h$^{-1}$ Mpc.  A selection
function is given by
\begin{equation}
 \bar{n}({\bf x}) = \frac{1}{1+a(x_3/L)^b},
\end{equation}
where $L$ is the size of the sample. To be comparable with, for
example, the LCRS selection function, the parameters are adopted
to be $L= 500$ h$^{-1}$ Mpc, $a =30$ and $b=3$.

Fig. 1 displays the DWT diagonal power spectrum $P_{jjj}$ of
particle distributions for three typical models with the selection
function eq.(14), in which we take $j_3=7$. It shows that the
power spectrum estimator eqs.(10) and (15) can perfectly recover
the DWT power spectrum regardless the selection functions (16).
This result will keep unchanged if $j_3\geq 7$. It means that
$n^{00j_3}({\bf x_3})$ gives a proper estimation of selection
function if $j_3$ is high enough. Practically, one can find a
properly recovered power spectrum by checking whether it is
insensitive to $j_3$.

It should be pointed out that with the developed algorithm the power 
spectrum eq.(8) is normalized scale-by-scale. The fluctuation 
amplitudes $\tilde{\epsilon}^g_{\bf j,l}$ (WFCs) on scale ${\bf j}$
is normalized by $n({\bf j,l})$ or $n^{(0,0,j_3)}(x_3)$ [eq.(15)],
which contains fluctuations on all scales larger than ${\bf j}$.
Therefore, the normalization factor generally is scale-dependent. 
This is different from conventional normalization, which is 
scale-independent.

If the field is Gaussian, there is no correlation between 
$\tilde{\epsilon}^g_{\bf j,l}$ and $n({\bf j,l})$ or $n^{(0,0,j_3)}(x_3)$.
Eq.(8) yields
\begin{equation}
P_j= \left \langle 
 \frac{1}{n^2({\bf j,l})} \right \rangle
  \left \langle [\tilde{\epsilon}^g_{\bf j,l}]^2\right \rangle -
  \left \langle \frac{1}{n({\bf j,l})}\right \rangle.
\end{equation}
It has been shown with the so-called ``partition of unity" of wavelets
that $\langle 1/n^2({\bf j,l})\rangle$ is approximately independent of 
${\bf j}$. In this case, the power spectrum eq.(8) is the same as that
given by conventional normalization (Jamkhedkar, Bi \& Fang 2001).

However, for non-Gaussian field, especially, if the perturbations on 
different scales are correlated, the small scale fluctuations given by the 
WFC $\tilde{\epsilon}^g_{\bf j,l}$ generally are correlated with large 
scale fluctuation contained in $n({\bf j,l})$ or $n^{(0,0,j_3)}(x_3)$.
In this case, the power spectrum eq.(8) is very different from usual 
power spectrum which doesn't not sensitive to the correlation between
perturbations on different scales. 

The power spectrum defined by eq.(8) will benefit to calculate the effect 
of selection function upon the redshift distortion (\S 3.3).

\section{Redshift distortion in the DWT representation}

\subsection{Velocity field}

The redshift distortion is due to peculiar motion of galaxies. For
a given mass field $\delta({\bf x})$, the galaxy velocity ${\bf
v}({\bf x})$ is a random field with mean
\begin{equation}
{\bf V}({\bf x})=\langle {\bf v}({\bf x})\rangle_v,
\end{equation}
where $\langle..\rangle_v$ denotes ensemble average for velocity
fields. ${\bf V}({\bf x})$ is the bulk velocity at ${\bf x}$.

In linear regime, the  bulk velocity is related to the density
contrast by
\begin{equation}
\delta({\bf x})= -\frac{1}{H_0\beta}\nabla\cdot {\bf V}({\bf x}),
\end{equation}
where $\beta \simeq \Omega_0^{0.6}/b$ is the redshift distortion
parameter at present, i.e. redshift $z\simeq 0$.

The {\it rms} deviation of velocity ${\bf v}({\bf x})$ from the
bulk velocity ${\bf V}({\bf x})$ is
\begin{equation}
\langle [v_i({\bf x})-V_i({\bf x})]^2\rangle_v
  =\sigma^2_{pv}({\bf x}).
\end{equation}
In the DWT representation, one can calculate $\sigma^2_{pv}({\bf
x})$ by the variance of the WFCs of velocity field, i.e.
$\tilde{\epsilon}^v_{\bf j,l}= \int {\bf v}({\bf x})\psi_{\bf
j,l}({\bf x})d{\bf x}$, which is actually the DWT pairwise
peculiar velocity (paper III).

In the paper III, we analyzed the PVD of the velocity fields
given by the N-body simulation for the CDM family of models. On
large scales, the velocity field is basically gaussian. It is
completely described by its mean ${\bf V}({\bf x})$ and variance
$\sigma_{pv}({\bf x})$. On small scales, the velocity field is 
significantly
non-gaussian. The one point function of $\tilde{\epsilon}^v_{\bf
j,l}$ is lognormal. The two-point pairwise velocity correlation
functions are non-zero on scales $\leq$ 1 h$^{-1}$ Mpc. Moreover,
on these scales, the pairwise velocities are also correlated
significantly with density fluctuations.

\subsection{The DWT power spectrum in redshift space}

The position of galaxy $i$ in redshift space is given by ${\bf
s_i}={\bf x_i} + \hat{\bf r}v_r({\bf x_i})/H_0$, where $v_r$ is
the radial component of ${\bf v}({\bf x})$. The number density
distribution in redshift space is then
\begin{equation}
n^S({\bf s})=
  \sum_{i=1}^{N_g}w_i\delta_D[{\bf s-x}_i- \hat{\bf r}v_r({\bf x}_i)/H_0]=
  \bar{n}^S({\bf s})[1+\delta^S({\bf s})],
\end{equation}
where $\bar{n}^S({\bf s})$ is the selection function in redshift
space. Similar to equation (8), the power spectrum in redshift
space is
\begin{equation}
P^S_{\bf j} = \frac{1}{2^{j_1+j_2+j_3}}\sum_{l_1=0}^{2^{j_1}-1}
 \sum_{l_2=0}^{2^{j_2}-1}\sum_{l_3=0}^{2^{j_3}-1}
  [\tilde\epsilon^S_{\bf j,l}]^2 -
\frac{1}{2^{j_1+j_2+j_3}}\sum_{l_1=0}^{2^{j_1}-1}
\sum_{l_2=0}^{2^{j_2}-1}\sum_{l_3=0}^{2^{j_3}-1}
\int\frac{\psi_{\bf j,l}^2({\bf x})}{\bar{n}^S({\bf x})}
 d{\bf x},
\end{equation}
where
\begin{equation}
 \tilde{\epsilon}^{S}_{\bf j, l}
   =\int \frac{n^S({\bf s})}{\bar{ n}^S({\bf s})}
    \psi_{\bf j,l}({\bf s})d{\bf s}.
\end{equation}
The selection function $\bar{n}^S({\bf s})$ can be determined from
the observed distribution $n^S({\bf s})$ by equation (13) or (14).
It can then be absorbed into the weight $w_i$ as equation (11) or
(15). The effect of the difference between $n^S({\bf s})$ and
$n({\bf s})$ will be studied in next section. In this section, we
will ignored this effect. Thus, using an auxiliary vector ${\bf
J}$, equations (21) and (23) yield
\begin{equation}
 \tilde{\epsilon}^{S}_{{\bf j,l}} =\sum_{i=1}^{N_g}w_i
\int d{\bf s} \left.
 \delta_D \left ({\bf s} -{\bf x}_i+ i \nabla_J \right )
 e^{i{\bf J}\cdot\hat{\bf r}v_r({\bf x}_i)/H_0}
  \psi_{\bf j,l}({\bf s})  \right |_{{\bf J}=0},
\end{equation}
where $\nabla_J$ is gradient operator on ${\bf J}$.

Subjecting equation (24) to an average over ensemble of velocity,
if the velocity field is {\it gaussian}, we have
\begin{eqnarray}
\langle  \tilde{\epsilon}^{S}_{{\bf j,l}}\rangle_v &  = &
\sum_{i=1}^{N_g}w_i \int d{\bf s} \left.
 \delta_D \left ({\bf s} -{\bf x}_i+ i \nabla_J \right )
 \langle e^{i{\bf J}\cdot\hat{\bf r}v_r({\bf x}_i)/H_0} \rangle_v
  \psi_{\bf j,l}({\bf s})  \right |_{{\bf J}=0} \\ \nonumber
  & = & \sum_{i=1}^{N_g}w_i
\int d{\bf s} \left.
 \delta_D \left ({\bf s} -{\bf x}_i+ i \nabla_J \right )
 e^{i{\bf J}\cdot\hat{\bf r}V_r({\bf x}_i)/H_0-
  (1/2)\sigma^2_{pv}({\bf x}_i)({\bf J}\cdot\hat{\bf r})^2H_0^2 }
  \psi_{\bf j,l}({\bf s})  \right |_{{\bf J}=0}.
\end{eqnarray}
For simplicity, we will use $\tilde{\epsilon}^{S}_{{\bf j,l}}$ for
$\langle  \tilde{\epsilon}^{S}_{{\bf j,l}} \rangle_v$ below
without causing confusion.

If we consider only linear effect of the bulk velocity, equation
(24) gives
\begin{eqnarray}
 \tilde{\epsilon}^{S}_{{\bf j,l}} & = & \sum_{i=1}^{N_g}w_i
\int d{\bf s} \left.
 \delta_D \left ({\bf s} -{\bf x}_i+ i \nabla_J \right )
\left  [1+ i\frac{1}{H_0}{\bf J}\cdot\hat{\bf r}V_r({\bf x}_i)
\right ] e^{-(1/2)\sigma^2_{pv}({\bf x}_i)({\bf J}\cdot\hat{\bf
r})^2 } \psi_{\bf j,l}({\bf s})  \right |_{{\bf J}=0} \\ \nonumber
  & = &
 \sum_{i=1}^{N_g}w_i
 \int d{\bf s}  \psi_{\bf j,l}({\bf s})
   e^{-(1/2)\sigma^2_{pv}({\bf s})(\hat{\bf r}\cdot \nabla )^2}
   \delta_D ({\bf s} -{\bf x}_i)  \\ \nonumber
    & - &
 \frac{1}{H_0}\sum_{i=1}^{N_g}w_i
 \int d{\bf s} \left.
 \hat{\bf r}
   \cdot [\nabla_s V_r({\bf s}+i\nabla_J)
    \delta_D \left ({\bf s} -{\bf x}_i+ i \nabla_J \right )]
   e^{-(1/2)\sigma^2_{pv}({\bf x}_i)({\bf J}\cdot\hat{\bf r})^2 }
   \psi_{\bf j,l}({\bf s})  \right |_{{\bf J}=0} \\ \nonumber
\end{eqnarray}
Neglecting the terms of the order of $V_r({\bf x})\delta({\bf
x})$, and using the linear relation eq.(19), equation (26) gives
\begin{eqnarray}
 \tilde{\epsilon}^{S}_{{\bf j,l}} &  = &
 \int d{\bf s} \psi_{\bf j,l}({\bf s})
  e^{(1/2)\sigma^2_{pv}({\bf s})(\hat{\bf r}\cdot \nabla )^2} n^g({\bf s})
      \\ \nonumber
  & + &
 \beta \int d{\bf s} \psi_{\bf j,l}({\bf s})
(\hat{\bf r} \cdot \nabla_s)^2 \nabla^{-2}
e^{-(1/2)\sigma^2_{pv}({\bf s})(\hat{\bf r}\cdot \nabla )^2}
n^g({\bf s})
\end{eqnarray}
Because all operators in the integrand of equation (27) are nearly
diagonal in the DWT representation (Farge et al 1996), equation
(27) can be rewritten as
\begin{equation}
  \tilde{\epsilon}^{S}_{{\bf j,l}}
 =(1+ \beta S_{\bf j})s^{pv}_{\bf j}\tilde{\epsilon}_{{\bf j,l}}
\end{equation}
where
\begin{equation}
 S_{\bf j} =  \int \psi_{\bf j,l}({\bf x})
  (\hat{\bf r} \cdot \nabla)^2\nabla^{-2}
    \psi_{\bf j,l}({\bf x}) d{\bf x}.
\end{equation}
and
\begin{equation}
 s^{pv}_{\bf j} = \int  \psi_{\bf j,l}({\bf x})
  e^{(1/2)\sigma^2_{pv}({\bf x})(\hat{\bf r}\cdot \nabla )^2}
   \psi_{\bf j,l}({\bf x}) d{\bf x}
\end{equation}
The method of calculating $S_{\bf j}$ and  $s^{pv}_{\bf j}$ is
presented in Appendix A.

Substituting equation (28) into equation (22), we have the
redshift distorted power spectrum as
\begin{equation}
 P^S_{\bf j}=(1+\beta S_{\bf j})^2 S^{PV}_{\bf j} P_{\bf j}.
\end{equation}
where $S^{PV}_{\bf j}=[s^{pv}_{\bf j}]^2$. Above equation
formulates the redshift distortion effect in DWT expression. This
is the basic formula for our redshift distortion analysis.
Usually, the factor $(1+ \beta S_{\bf j})^2$ is called linear
redshift distortion, and $S^{PV}_{\bf j}$ called non-linear
redshift distortion due to the pairwise velocity dispersion.
However in our derivation, the two parts are not treated
separately.

This derivation can be generalized to any velocity fields, which
are not simply described by equations (18) and (20). For
instance, if the pairwise velocities are correlated, i.e.
\begin{equation}
\langle [v_i({\bf x})-V_i({\bf x})][v_i({\bf x'})-V_i({\bf
x'})]\rangle_v =\sigma^2_{pv}({\bf x-x'}),
\end{equation}
the pairwise velocity dispersion factor becomes
\begin{eqnarray}
 \lefteqn{ S^{PV}_{\bf j}  = \int \int d{\bf x}d{\bf x'} } \\ \nonumber
  & & \psi_{\bf j,l}({\bf x})\psi_{\bf j,l}({\bf x'})
 e^{[(1/2)\sigma^2_{pv}({\bf x})(\hat{\bf r}\cdot \nabla )^2+
    \sigma^2_{pv}({\bf x-x'})(\hat{\bf r}\cdot \nabla )
      (\hat{\bf r}\cdot \nabla' )
 +(1/2)\sigma^2_{pv}({\bf x'})(\hat{\bf r}\cdot \nabla' )^2]}
   \psi_{\bf j,l}({\bf x}) \psi_{\bf j,l}({\bf x'}),
\end{eqnarray}
where $\nabla'$ is gradient operator on ${\bf x'}$.

\subsection{Effect of selection functions}

The theory of redshift distortion presented in last section did
not consider the effects of selection functions, $\bar{n}^S({\bf
s})$ and $\bar{n}({\bf s})$. Since selection function relies on
the radial distance, it could be also a source of the anisotropy
of power spectrum with respect to redshift direction. It should be
taken into account when analyze observed samples.

In the linear approximation of $v_r$, equation (21) gives
\begin{equation}
 n^S({\bf s})=n^g({\bf s})-
  \sum_{i=1}^{N_g}\frac{1}{H_0}v_r({\bf x_i})
     \hat{\bf r}\cdot \nabla\delta_D({\bf s-x}_i)
   =n^g({\bf s}) -\frac{1}{H_0} \hat{\bf r}\cdot \nabla
   [n^g({\bf s})V_r({\bf s})].
\end{equation}
In the second steps, $v_r$ is replaced by $V_r$, as only the bulk
velocity is considered. Using equations (1) and (21), equation
(34) yields
\begin{equation}
 \delta^S({\bf s}) \simeq -1+\frac{\bar{n}({\bf s})}{\bar{n}^S({\bf s})}
   +\frac{\bar{n}({\bf s})}{\bar{n}^S({\bf s})}
   \left \{ \delta({\bf s}) - \frac{1}{H_0\bar{n}({\bf s})}
    \hat{\bf r}\cdot \nabla [\bar{n}({\bf s})V_r({\bf s})] \right \},
\end{equation}
where the second order term $\delta({\bf s})V_r({\bf s})$ is
ignored.

The term -1 on the r.h.s. of equation (35) does not contribute to
power spectrum, because of $\int\psi_{\bf j,l}(x)dx=0$. In the
linear approximation, $\bar{n}({\bf s})/\bar{n}^S({\bf s})\simeq
1 + O(v_r)$ and therefore, the factor $\bar{n}^g({\bf
s})/\bar{n}^S({\bf s})$ in the third term of the r.h.s. of
equation (34) can be approximated as 1.

The second term in the r.h.s. of equation (35) contains a linear
term of $V_r$, i.e. the same order as the third term. However,
with the selection function equation (13), we have
\begin{equation}
 \int \bar{n}^g({\bf x})
  \psi_{\bf j,l}({\bf x})d{\bf x}=\int \bar{n}^{({\bf j'})}({\bf x})
     \psi_{\bf j,l}({\bf x})d{\bf x}=0
\hspace{3mm}
 {\rm if\ one\ of}\ j_i\ (i=1,2,3)\ {\rm satisfies} \ \  j_i \geq j'_i.
\end{equation}
For plane parallel approximation, ${\bf j'}=(0,0,j'_3)$, the
above equation is always hold if we study fluctuation powers on
scales $j_1
>0$ or $j_2 >0$. Actually, in this case,
$\bar{n}^g({\bf x})$ or  $\bar{n}^{({\bf j'})}({\bf x})$ depend
only on $x_3$, and thus their projections onto bases
$\psi_{j,l}(x_1)$ or $\psi_{j,l}(x_2)$ ($j>0$) are always null.
Similarly the selection function in redshift space is also only
dependent on $x_3$. Accordingly, in the plane parallel
approximation, we have
\begin{equation}
 \int \frac{\bar{n}({\bf s})}{\bar{n}^S({\bf s})}\psi_{\bf j,l}d{\bf s}
  =0
\hspace{3mm} {\rm if} j_1\ {\rm or}\ j_2 >0.
\end{equation}
which implies that the second term in the r.h.s. of equation (34)
also has not contribution to the projection on bases $\psi_{\bf
j,l}(x_1)$ or $\psi_{\bf j,l}(x_2)$.

In fact, the condition of plane parallel approximation is not
necessary. For radial redshift, the selection function is still a
1-D function. Its projection onto the bases $\psi_{\bf j,l}$ on
the celestial spherical surface is still zero.  Thus, the linear
redshift distortion mapping is given by
\begin{equation}
 \delta^S({\bf s}) \simeq
    \delta({\bf s}) - \frac{1}{H_0\bar{n}({\bf s})}
    \hat{r}\cdot \nabla [\bar{n}({\bf s})v_r({\bf s})].
\end{equation}
Hereafter we will use ${\bf x}$ for the variable ${\bf s}$. It
will not cause confusion as the superscript $S$ stands for
redshift space.

In the plane parallel approximation, equation (38) becomes
\begin{equation}
 \delta^S({\bf x})=\delta({\bf x}) -
     \frac{1}{H_0}\frac{\partial v_3}{\partial x_3}
   - \frac{1}{H_0}\frac{d\ln \bar{n}(x_3)}{dx_3}v_3({\bf x}),
\end{equation}
Using equation (19), $v_3$ can be represented by $\delta({\bf
x})$, we have then
\begin{equation}
 \delta^S({\bf x})= \left [1 +
     \beta\frac{\partial^2}{\partial x_3^2}\nabla^{-2}
   + \beta\frac{d\ln \bar{n}(x_3)}{dx_3}
     \frac{\partial}{\partial x_3}\nabla^{-2} \right ] \delta({\bf x}).
\end{equation}
The differential operator of the second term in the bracket of
equation (40) is nearly diagonal in the DWT representation (Farge
et al 1996.) Thus, we have
\begin{equation}
 \tilde{\epsilon}_{\ \ \bf j,l}^{S} \simeq
 (1+ \beta S_{\bf j}) \tilde{\epsilon}_{\bf j,l}+
  \left . \beta\frac{d\ln \bar{n}(x_3)}{dx_3}\right|_{\bf j,l}
    \sum_{l_3-l'_3} Q_{{\bf j}, l_3-l'_3}\tilde{\epsilon}_{{\bf
j},l_1,l_2,l'_3}
\end{equation}
where $d\ln \bar{n}^S(x_3)/dx_3|_{\bf j,l}$ means the value of
$d\ln \bar{n}^S(x_3)/dx_3$ in the mode ${\bf j,l}$. The
coefficient $Q_{{\bf j},l_3-l'_3}$ are defined by
\begin{equation}
 Q_{{\bf j}, l_3-l'_3}=\int \psi_{{\bf j},l_1,l_2,l_3}({\bf x})
 \frac{\partial}{\partial x_3}\nabla^{-2}\psi_{{\bf j},l_1,l_2,l'_3}
 ({\bf x}) d{\bf x}.
\end{equation}
The calculations of $Q_{{\bf j},l_3-l'_3}$ are given in Appendix
A.

Since $Q_{{\bf j}, 0}=0$ (Appendix A), the first and second terms
on the r.h.s. of equation (40) are not correlated, we have
\begin{equation}
 \langle |\tilde{\epsilon}_{\ \ \bf j,l}^{S}|^2\rangle =
 [1+ \beta S_{\bf j}]^2
    \langle |\tilde{\epsilon}_{\bf j,l}|^2  \rangle +
  \left [\left . \beta\frac{d\ln \bar{n}(x_3)}{dx_3}\right|_{\bf
j,l}\right ]^2
    \sum_{l_3-l'_3} Q^2_{{\bf j}, l_3-l'_3}
      \langle |\tilde{\epsilon}_{{\bf j},l_1,l_2,l'_3}|^2  \rangle
\end{equation}
For uniform fields, $\langle |\tilde{\epsilon}_{\ \ \bf
j,l}^{S}|^2\rangle$ and $\langle |\tilde{\epsilon}_{\bf j,l}|^2
\rangle$ are ${\bf l}$-independent. Hence, equation (43) gives the
relation between the DWT power spectra in redshift $P^S_{\bf j}$
and real spaces $P_{\bf j}$ as
\begin{equation}
 P^S_{\bf j}= \left \{[1+ \beta S_{\bf j}]^2+
 \left [\left . \beta\frac{d\ln \bar{n}(x_3)}{dx_3}\right|_{\bf
j,l}\right ]^2
 \sum_{l_3-l'_3} Q^2_{{\bf j}, l_3-l'_3} \right \}P_{\bf j}.
\end{equation}
which quantifies both redshift distortion and selection effect on
the DWT power spectrum.

Using inequality equation (A15), we can show that if
\begin{equation}
 \frac{d\ln \bar{n}(x_3)}{dx_3} <2\pi n_p \frac{2^{j_3}}{L_3},
\end{equation}
we have
\begin{equation}
 \left [ \left . \beta \frac{d\ln \bar{n}(x_3)}{dx_3}\right|_{\bf
j,l}\right ]^2
 \sum_{l_3-l'_3} Q^2_{{\bf j}, l_3-l'_3} < \beta^2 S^2_{\bf j}.
\end{equation}
which states that the selection function term in equation (44) is
even less than the second order terms $\beta^2 S^2_{\bf j}$ if the
selection function varies with $x_3$ slowly. In this case, the
selection function does not significantly disturb the linear
redshift distortion described by equation (31) when we perform a
DWT power spectrum analysis using equations (10) and (15).

The Poisson noise term in equation (10) is not affected by the
redshift distortion, i.e. it is the same as the Poisson noise
term in equation (22). Because both Poisson noise terms of
equations (10) and (22) linearly depend on $n({\bf x})$ or
$n^S({\bf x})$, for an ensemble average, we have $\langle n({\bf
x}) \rangle = \langle n^S({\bf x}) \rangle $.

\section{Simulation samples}

To demonstrate the redshift distortion in the DWT representation,
we use the N-body simulation samples like that in Paper II. The
model parameters used are listed in table. 1. We use modified
AP$^3$M code (Couchman, 1991) to evolve $128^3$ cold dark matter
particles in a periodic cube of side length $L$. The linear power
spectrum is using the fitting formula given in Bardeen et al.
(1986).

\begin{table*}
 \begin{center}
 \centerline{Table 1}
 \bigskip
 \begin{tabular}{cccccclc}
 \hline\hline
   Model& L/h$^{-1}$Mpc & $\Omega_0$ & $\Lambda$ &
     $\Gamma$ & $\sigma_8$ & $\beta$ & realizations\\
   \hline
  SCDM          & 256 & 1.0 & 0.0 & 0.5  & 0.55 & 1.0 & 10 \\
  $\tau$CDM     & 256 & 1.0 & 0.0 & 0.25 & 0.55 & 1.0 & 10 \\
  $\Lambda$CDM1 & 256 & 0.3 & 0.7 & 0.21 & 0.85 & 0.49& 10 \\
  $\Lambda$CDM2 & 480 & 0.3 & 0.7 & 0.21 & 0.95 & 0.49& 6 \\\hline
 \end{tabular}
 \end{center}
\end{table*}

In our simulation, we use the so-called ``glass" configuration to
generate the unperturbed uniform distribution of particles, and
the Zel'dovich approximation to set up the initial perturbation.
The triangular-shaped cloud (TSC) method is used for the mass
assignment on the grid and the calculation of the force on a
given particle from interpolation of the grid values. We take 600
total integration steps from $z_i=15$ for the SCDM model, and
$z_i=25$ for $\Lambda$CDM and $\tau$CDM down to $z=0$. The force
softening parameter $\eta $ in the comoving system decreases with
time as $\eta \propto 1/a(t)$. Its initial value is taken to be
15\%, and the minimum value to be 5\% of the grid size,
respectively.

\section{$\beta$-estimators and reconstruction}

\subsection{A test of the linear approximation}

We try to estimate the redshift distortion parameter $\beta$ using
equation (30). Obviously, the precision of the $\beta$
determination is dependent on the linear relation of equation
(18), which is valid only on the scales where the bulk velocity
can be described by linear or quasi-linear density perturbations.
It has been already realized that the non-linearity of the
relation between $\delta({\bf x})$ and ${\bf V}({\bf x})$ will be
significant on small scales (e.g. Kudlicki et al. 2000).
Therefore, it is necessary to have an estimation of uncertainties
due to the non-linear $\delta({\bf x})-\bf{V}({\bf x})$ relation.

The non-linear effect can be estimated by a ``blueshift" (change
the sign of velocity from ${\bf v}({\bf x})$ to $-{\bf v}({\bf
x})$) distorted power spectrum. From the derivation of equation
(31) in last section, it is easy to show that, the blueshifted
power spectrum is
\begin{equation}
 P^B_{\bf j}=(1-\beta S_{\bf j})^2 S^{PV}_{\bf j} P_{\bf j}.
\end{equation}
Which has the same non-linear redshift distortion effect
$S^{PV}_{\bf j}$ as the redshift power spectrum. The difference
between the redshift [equation (31)] and blueshift [equation (47)]
distorted power spectra is a sign of the linear term $\beta$.
Combining equations (31) and (47), one can determine $\beta$ as
\begin{equation}
 \beta =\frac{1}{S_{\bf j}}\frac {(\sqrt{P^S_{\bf
 j}}-\sqrt{P^B_{\bf j}})}
    {(\sqrt{P^S_{\bf j}}+\sqrt{P^B_{\bf j}})}.
\end{equation}

We test equation (48) by simulation samples in the $\Lambda$CDM model. 
First, we calculate the diagonal power spectra of particles, 
$P^S_{jjj}$ and $P^B_{jjj}$, in the plane parallel approximation. 
We then find $\beta$ by equation (48). The result is presented in 
Fig. 2. The density parameter used for the simulation is 
$\Omega=0.3$, or $\beta = 0.49$. Fig. 2 shows that the values of 
$\beta$ given by equation (48) is generally overestimated. But the 
overestimations are no more than 10\% on scales $k < 0.5$ h Mpc$^{-1}$, 
and about 20\% for $k > 1$ h Mpc$^{-1}$. It should be pointed out that 
the goodness of the resulted $\beta$ on small scales $k > 1$ h Mpc$^{-1}$
doesn't mean that the linear relation eq.(19) can be used to describe 
the redshift distortion on these scales. The non-linear effect on these 
scales can not be ignored, and the redshift distortion is dominated by
the term $S_{\bf j}^{PV}$. The goodness for $k > 1$ h Mpc$^{-1}$ shown 
in Fig. 2 is due to that the non-linear effect is significantly 
repressed by the test eq.(48) based on an assumed blueshifts.  

\subsection{Symmetry of the DWT quantities}

At the first glance, it seems impossible to estimate $\beta$
using the observed redshift distorted power spectrum $P^S_{\bf j}$
only, as all the quantities such as $\beta$, $P_{\bf j}$, and
$S^{PV}_{\bf j}$ (or $\sigma_{pv}$) included on the r.h.s. of
equation (30) are unknown. Usually, equation (30) is used to fit
the observed power spectrum with model-predicted $P_{\bf j}$, and
then determine the parameter $\beta$ from the most-likely-fitting.

But we try to search for $\beta$-estimators which depend on model
as less as possible. To achieve this, we take the advantage of the
wavelet analysis: the DWT modes of $\psi_{j_1,j_2,j_3}({\bf x})$
are not rotational invariant, but cyclic permutational invariant.
As a consequence, all the DWT quantities in the redshift
distortion equation (30), $P_{\bf j}$, $P^S_{\bf j}$, $S_{\bf j}$,
and $S^{PV}_{\bf j}$, are dependent on the three indexes
$(j_1,j_2,j_3)$, rather than the length scale of mode ${\bf j}$
only. The quantities $P_{\bf j}$, $P^S_{\bf j}$, $S_{\bf j}$ and
$S^{PV}_{\bf j}$ satisfy the following symmetry.

1. If cosmic density and velocity fields are statistically
isotropic, the DWT power spectrum in real space is invariant with
respect to the cyclic permutations of index  ${\bf
j}=(j_1,j_2,j_3)$, i.e.
\begin{equation}
 P_{j_1,j_2,j_3}=P_{j_3,j_1,j_2}=P_{j_2,j_3,j_1}
\end{equation}

2. In the plane-parallel approximation, e.g., coordinate $x_3$ is
in the redshift direction, we have
\begin{equation}
 P^S_{j_1,j_2,j_3}=P^S_{j_2,j_1,j_3}
\end{equation}
\begin{equation}
 S_{j_1,j_2,j_3}=S_{j_2,j_1,j_3}
\end{equation}
\begin{equation}
 S^{PV}_{j_1,j_2,j_3}=S^{PV}_{j_2,j_1,j_3}
\end{equation}

3. If $\sigma_{pv}$ is constant, i.e. scale-independent,
following equation (A10), we have
\begin{equation}
 S^{PV}_{j_1,j_2,j_3} =S^{PV}_{j_3},
\end{equation}
i.e. $S^{PV}_{\bf j}$ is independent of $j_1$ and $j_2$. More
generally, equation (53) also hold even when the radial
correlation of pairwise peculiar velocity is considered [equation
(33)].

Equations (49) - (53) provide the base of designing the
$\beta$-estimators with the DWT power spectrum.

\subsection{$\beta$-estimator with scale-independent $\sigma_{pv}$}

Assuming that the pairwise velocity dispersion is
scale-independent, equations (31) and (49) give
\begin{equation}
\frac{P^S_{jj_2j_3}}{P^S_{ jj_3j_2}}\simeq
   \frac{(1+\beta S_{ jj_2j_3})^2}{(1+\beta S_{ jj_2j_3})^2}
    \left [\frac{S^{PV}_{jj_2j_3}}{S^{PV}_{jj_3j_2}} \right ]^2.
\end{equation}
For a given pair $(j_2,j_3)$, $S^{PV}_{jj_2j_3}/S^{PV}_{jj_3j_2}$
is a constant, and thus the r.h.s. of equation (54) depends only
on two parameters $\beta$ and
$S^{PV}_{jj_2j_3}/S^{PV}_{jj_3j_2}$. Accordingly, these parameters
can be found by the best fitting of the r.h.s. of equation (54)
with observed ratios $P^S_{jj_2j_3}/P^S_{ jj_3j_2}$, $j=2,3..$.

A numerical example of this fitting is demonstrated in Fig. 3, in
which we take $(j_2,j_3)=(2,3)$, and $j=2...7$. We calculate
$P^S_{jj_2j_3}/P^S_{ jj_3j_2}$ for the $\Lambda$CDM simulation
samples, and the best fitting yields the values of $\beta=0.53\pm
0.25$ and $S^{PV}_{j23}/S^{PV}_{j32}=0.80\pm 0.04$. The precision
of the estimator eq.(53) probably is not better than 20\%. This
shows that the assumption of a constant $\sigma_{pv}$ is not too
bad, but its effect on the $\beta$-determination cannot be
neglected.

\subsection{$\beta$-Estimator with scale-dependent $\sigma_{pv}$}

If $\sigma_{pv}$ is scale-dependent, equation (54) is no longer
correct in general. In this case $S^{PV}_{j_1,j_2,j_3}$ depends
on the three index $(j_1,j_2,j_3)$, rather than $j_3$ only. So
the relation eq.(30) is obviously not enough to extract $\beta$
from measuring redshift distorted power spectrum $P^S_{\bf j}$
only. To search for an appropriate algorithm for $\beta$
estimation, we shall first consider the property of $S^{PV}_{\bf j}$.

At first, the pairwise velocity dispersion factor $S^{PV}_{\bf j}$
is determined by an isotropic function $\sigma_{pv}({\bf x})$,
which is scale-dependent. Let us consider the modes of
$L/2^{j_\perp} <L/2^{j_3}$, where $j_\perp$ is defined by
$2^{2j_{\perp}} \equiv 2^{2j_1}+2^{2j_2}$. In this case, the
scales of these modes are dominant by $j_3$, and therefore,
$\sigma_{pv}({\bf x})$ is also dominant by $j_3$. That is, for a
given $j_3$, $S^{PV}_{\bf j}$ will keep constant if $j_\perp >
j_3$.

To test this expectation, we calculate $S^{PV}_{\bf j}$ in the
$\Lambda$CDM model. The result is shown in Fig. 4.  One can see
from Fig. 4 that $S^{PV}_{\bf j}$ is generally dependent on
indexes $j_\perp$ as well as $j_3$. However, for a given $j_3$,
$S^{PV}_{\bf j}$ almost keeps constant in the range of
$j_{\perp}>j_3$.  Since $j=2,3$ is the two largest scales of the
samples, $S^{PV}_{j_1j_2 2}$ and $S^{PV}_{j_1j_2 3}$ keep constant
very well for all $j_1,j_2$.

Using this result, we can design a $\beta$-estimator as follows
\begin{equation}
 \frac{P^S_{j23}P^S_{j'32}}{P^S_{j'23}P^S_{j32}}
   =\frac{(1+\beta S_{j23})^2(1+\beta S_{j'32})^2}
      {(1+\beta S_{j'23})^2(1+\beta S_{j32})^2},
\end{equation}
or
\begin{equation}
 \beta \simeq \left [\left(
 \frac{P^S_{j23}P^S_{j'32}}{P^S_{j'23}P^S_{j32}}
   \right)^{1/2} -1\right ]
 \frac{1}{(S_{j23}-S_{j32}+S_{j'32}-S_{j'23})}
\end{equation}
Obviously, the first factor on the r.h.s. of equation (56) is to
measure the difference between spectra with the same scales, but
different shapes of the modes. The estimator eq.(56) is
model-free, as equation (56) is based only on property that two
DWT modes of $(j_{\perp}, j_3)$ and $(j'_{\perp}, j_3)$ have the
same scale, but different shapes if $j_{\perp}>j_3$ and
$j'_{\perp}>j_3$.

To apply the estimator (56), we take $j=2$ and $j'=7$, because
modes with $j=2$ and $j'=7$ have largest difference in the shape,
but very small difference in the scale. This estimator yields the
values of $\beta=0.47\pm 0.18$ for $\Lambda$CDM sample with
simulation box $L=$ 480 h$^{-1}$ Mpc; $0.93\pm 0.22$ for SCDM
sample with $L=$ 256 h$^{-1}$ Mpc; and $1.00\pm 0.34$ for
$\tau$CDM sample with $L=$ 256 h$^{-1}$ Mpc. Considering the
uncertainty of non-linearity is about 20\% (\S 5.1), the
estimator eq. (56) works very well.

\subsection{Estimation of $\sigma_{pv}$}

After the $\beta$ estimation, we can also estimate $\sigma_{pv}$
on scale $j_3$ using off-diagonal power spectrum in redshift
space. From \S 5.4, we know that $S^{PV}_{\bf j}$ is almost
constant in the range of $j_{\perp}>j_3$. So we have
$S^{PV}_{77j}=S^{PV}_{jjj}$ and $S^{PV}_{j77}=S^{PV}_{777}$.

We can calculate the ratio of pairwise velocity dispersion factor
$S^{PV}_{jjj}/S^{PV}_{777}$ by
\begin{equation}
\frac{P^S_{77j}(1+\beta S_{j77})^2}{P^S_ {j77}(1+\beta
 S_{77j})^2}=\frac{S^{PV}_{77j}}{S^{PV}_{j77}}=
 \frac{S^{PV}_{jjj}}{S^{PV}_{777}}.
\end{equation}
On the largest scale ${j}=2$, the nonlinear redshift distortion
due to the pairwise velocity dispersion is negligible, we have
$S^{PV}_{222}\simeq 1$. Thus, we have all the diagonal members of
the pairwise velocity dispersion factor $S^{PV}_{jjj}$.  The
parameter $\sigma_{pv}$ on the scale ${j}$ can then be found by
$S^{PV}_{jjj}=[s^{pv}_{jjj}]^2$ with equation (A10), i.e.
\begin{equation}
 s^{pv}_{j_1,j_2,j_3} =\frac{1}{2^{j_3}}
    \sum_{n_3 = \infty}^{\infty}
  |\hat{\psi}(n_3/2^{j_3})|^2 \exp[-(1/2)\sigma^2_{pv} (2\pi n_3/L_3)^2].
\end{equation}

For the $\Lambda$CDM simulation sample, $S^{PV}_{jjj}$ is shown
in Fig.5. The $\sigma_{pv}$ on scales ${j}=3...7$ is shown in Fig.
6. Although the values of $\sigma_{pv}$ shown in Fig. 6 are
correct in average, but it does not match with the direct
measurement of $\sigma_{pv}$ given in Paper III. Especially, on
small scales $j=$ 6 and 7, the values of $\sigma_{pv}$ are
significantly lower than the direct measurement. This is not
unexpected. The factor $s^{pv}_{\bf j}$ equation (58) is obtained
under the assumption of gaussian velocity field (\S 3.1). However,
we have shown in Paper III that the velocity field is actually
intermittent on small scales. The PDF of
$\tilde{\epsilon}^{v}_{\bf j,l}$ is not gaussian, but lognormal.

Nevertheless, the factor $S^{PV}_{\bf j}$ is still good for the
redshift-real mapping and $\beta$-estimation, because in these
calculations, we used only the symmetric properties and
scale-dependence of $S^{PV}_{\bf j}$, but not the details of
$\sigma_{pv}$. This point can also be seen from reconstruction of
power spectrum in real space from that in redshift space. Using
the $\beta$ estimated by equation (56), and $S^{PV}_{jjj}$ by
equation (57), one can reconstruct the diagonal DWT power
spectrum in real space $P_{jjj}$ from $P^S_{jjj}$ through
equation (31). Fig.(7) compares the recovery of $P_{jjj}$ with
the original diagonal DWT power spectrum in the $\Lambda$CDM
model. The result shows that the algorithm of reconstruction is
reliable.

\section{Conclusion}

We established the mapping between the DWT power spectra in real
and redshift spaces. From equations (31) and (44), the mapping in
the plane parallel approximation is
\begin{equation}
 P^S_{\bf j}= \left \{[1+ \beta S_{\bf j}]^2+
 \left [\left . \beta\frac{d\ln \bar{n}(x_3)}{dx_3}\right|_{\bf j,l}\right
 ]^2
 \sum_{l_3-l'_3} Q^2_{{\bf j}, l_3-l'_3} \right \}S^{PV}_{\bf j}
    P_{\bf j}.
\end{equation}
which includes the effects of (1) bulk velocity (the term of
$S_{\bf j}$), and (2) selection function (the term of
$\bar{n}(x_3)$) as well as (3) pairwise peculiar velocity (the
term of $S^{PV}_{\bf j}$).

In the Fourier representation, the redshift distortion mapping is
axially symmetric with respect to the redshift direction. This is
because the Fourier mode is rotational invariant, and the redshift
distortion produces an anisotropy in the spatial distribution of
galaxies between the line of sight and directions other than it.
An isotropic statistics plus an axially symmetric violation will
result in an axial symmetry. On the other hand, all redshift
distortion factors in equation (59) are no longer axial
symmetric, which is due to the modes ${\bf j}$ being not
invariant under rotational transformation, but for cyclical
permutation. Therefore, the symmetry of the DWT redshift
distortion results from a cyclical permutation plus an axially
symmetric violation.

By virtue of this feature, we develop $\beta$ estimators, which
are mainly based on the shape-dependence of the redshift distorted
DWT power spectrum, but not the scale-dependence. The estimators 
eqs.(54) and (56) look similar to the quadrupole-to-monopole ratio 
method based on the Fourier power spectrum in redshift space, which 
gives a model-free estimation of $\beta$ (e.g. Cole et
al, 1994, Fisher et al, 1994, Hatton \& Cole, 1998, Szalay et al,
1998). However, the estimators 
eqs.(54) and (56) is different from the quadrupole-to-monopole ratio
method. The latter considered only the linear redshift distortion
effect, but not the non-linear redshift distortion effect. If considering
the non-linear redshift distortion effect, the quadrupole-to-monopole 
ratio method needs a fitting of observed redshift distorted power 
spectrum with model-predicted power spectrum. On the other hand, 
the estimators eqs.(54) and (56) considered both linear 
($\beta S_{\bf j}$) and non-linear ($S^{PV}_{\bf j}$) redshift distortion 
effects. These $\beta$-estimators are model-free even when the 
non-linear redshift distortion effect is not negligible. We test the
$\beta$-estimators using N-body simulation samples. The result
shows that regardless the pairwise velocity dispersion is
scale-dependent or not, the $\beta$-estimators can yield the
correct number of $\beta$ with about 20\% uncertainty. We also
develop an algorithm for reconstruction of the power spectrum in
real space from the redshift distorted power spectrum. The
numerical tests also show that the real power spectrum can be
reasonably recovered from the redshift distorted power spectrum.

\acknowledgments

LLF and YQC acknowledges support from the National Science
Foundation of China (NSFC) and National Key Basic Research Science
Foundation.

\appendix

\section{Calculations of $S_{\bf j}$, $s^{pv}_{\bf j}$, and $Q_{\bf
j,l-l'}$}

\subsection{$S_{\bf j}$}

Let us consider plane-parallel approximation, i.e. coordinate
$x_3$ is in the redshift ($z$)-direction. The linear redshift
distortion $S_{\bf j}$ equation (28) gives
\begin{equation}
S_{\bf j} = \int \psi_{\bf j,l}({\bf x})
\frac{\partial^2}{\partial x_3^2}\nabla^{-2}\psi_{\bf j,l}({\bf
x}) d{\bf x}.
\end{equation}
Because 1-D wavelet $\psi_{j,l}(x)$ is given by dilating and
translating the basic wavelet $\psi(\eta)$ as
\begin{equation}
 \psi_{j,l}(x)=\left (\frac{2^j}{L}\right )^{1/2} \psi(\frac{2^jx}{L}-l),
\end{equation}

The Fourier transform of $\psi_{j,l}$ is
\begin{equation}
 \psi_{j,l}=\frac{1}{L}\sum_{n=-\infty}^{\infty}
 \hat{\psi}_{j,l}(n)e^{-i2\pi n x/L}
\end{equation}
and
\begin{equation}
 \hat{\psi}_{j,l}(n)=\left (\frac{L}{2^j}\right )^{1/2}
     \hat{\psi}(n/2^j)e^{-i2\pi n l/2^j},
\end{equation}
where $\hat{\psi}(n)$ is the Fourier transform of the basic
wavelet
\begin{equation}
 \hat{\psi}(n)=\int_{-\infty}^{\infty} \psi(\eta)e^{-i2\pi n\eta}d\eta.
\end{equation}
The function $|\hat{\psi}(n)|^2$ is shown in Fig. 1 of Yang et al.
(2001).

Thus, equation (A1) becomes
\begin{eqnarray}
 \lefteqn{ S_{j_1,j_2,j_3}= \frac{1}{2^{j_1+j_2+j_3}} } \\ \nonumber
   & & \sum_{n_1,n_2,n_3 = \infty}^{\infty}
   \frac{(n_3/L_3)^2}{(n_1/L_1)^2 + (n_2/L_2)^2+ (n_3/L_3)^2}
   |\hat{\psi}(n_1/2^{j_1})\hat{\psi}(n_2/2^{j_2})
   \hat{\psi}(n_3/2^{j_3})|^2.
\end{eqnarray}
Since $\hat{\psi}(n)$ is non-zero only around two peaks at $n=\pm
n_p$, the summation of equation (A6) actually only over few
number of $n_i$ around $\pm n_p 2^{j_i}$.

If $L_1=L_2=L_3=L$, equation (A6) becomes
\begin{equation}
  S_{j_1,j_2,j_3}= \frac{1}{2^{j_1+j_2+j_3}}
    \sum_{n_1,n_2,n_3 = \infty}^{\infty}
   \frac{n_3^2}{n_1^2 + n_2^2+ n_3^2}
   |\hat{\psi}(n_1/2^{j_1})\hat{\psi}(n_2/2^{j_2})
    \hat{\psi}(n_3/2^{j_3})|^2.
\end{equation}
For diagonal modes, i.e. $j_1=j_2=j_3=j$ we have
\begin{equation}
S_{j,j,j}=\frac{1}{3}
\end{equation}

\subsection{$s^{pv}_{\bf j}$}

If $\sigma_v({\bf x})$ is independent of ${\bf x}$, $s^{pv}_{\bf
j}$ [equation (28)] can be calculated by
\begin{equation}
 s^{pv}_{j_1,j_2,j_3} = \frac{1}{2^{j_1+j_2+j_3}}
    \sum_{n_1,n_2,n_3 = \infty}^{\infty}
|\hat{\psi}(n_1/2^{j_1})\hat{\psi}(n_2/2^{j_2})\hat{\psi}(n_3/2^{j_3})|^2
     \exp[-(1/2)\sigma^2_{pv} (\hat{r}\cdot {\bf n})^2],
\end{equation}
where vector ${\bf n}=2\pi(n_1/L_1, n_2/L_2, n_3/L_3)$. In
plane-parallel approximation, we have
\begin{equation}
s^{pv}_{j_1,j_2,j_3} = \frac{1}{2^{j_3}}
   \sum_{n_3 = \infty}^{\infty}
 |\hat{\psi}(n_3/2^{j_3})|^2 \exp[-(1/2)\sigma^2_{pv} (2\pi n_3/L_3)^2].
\end{equation}
The summation of equations. (A9) and (A10) also runs only over
few number of $n_i$ around $\pm n_p 2^{j_i}$.

\subsection{$Q_{\bf j,l-l'}$}

Similarly, for $Q_{\bf j,l-l'}$ given by [equation (41)]
\begin{equation}
 Q_{{\bf j}, l_3-l'_3}=\int \psi_{{\bf j},l_1,l_2,l_3}({\bf x})
 \frac{\partial}{\partial x_3}\nabla^{-2}
  \psi_{{\bf j},l_1,l_2,l'_3}({\bf x}) d{\bf x},
\end{equation}
we have
\begin{eqnarray}
 \lefteqn { Q_{{\bf j}, l_3-l'_3}=\frac{1}{2^{j_1+j_2+j_3}} } \\ \nonumber
   & & \sum_{n_1,n_2,n_3 = \infty}^{\infty}
 \frac{(n_3/L_3) \sin[2\pi n_3(l_3-l_3')/2^{j_3}]}
   {2\pi [(n_1/L_1)^2 + (n_2/L_2)^2+ (n_3/L_3)^2]}
|\hat{\psi}(n_1/2^{j_1})\hat{\psi}(n_2/2^{j_2})\hat{\psi}(n_3/2^{j_3})|^2.
\end{eqnarray}
Equation.(A12) gives
\begin{equation}
  Q_{{\bf j}, l_3-l'_3}=0, \ \ \ {\rm if} \ \  l_3-l'_3=0.
\end{equation}

 From equation (A12), we have
\begin{eqnarray}
 \lefteqn { Q_{\bf j, l-l'} = \frac{1}{2^{j_1+j_2+j_3}} }
   \\ \nonumber
   & & \sum_{n_1,n_2,n_3 = \infty}^{\infty}\frac{L_3}{n_3}
  \frac{(n_3/L_3)^2\sin[2\pi n_3(l_3-l_3')/2^{j_3}] }
   {2\pi [(n_1/L_1)^2 + (n_2/L_2)^2+ (n_3/L_3)^2]}
|\hat{\psi}(n_1/2^{j_1})\hat{\psi}(n_2/2^{j_2})\hat{\psi}(n_3/2^{j_3})|^2.
    \\ \nonumber
  & & \simeq \frac{L_3}{2\pi n_p2^{j_3}} \frac{1}{2^{j_1+j_2+j_3}}
    \\ \nonumber
  & &      \sum_{n_1,n_2,n_3 = \infty}^{\infty}
  \frac{(n_3/L_3)^2\sin[2\pi n_3(l_3-l_3')/2^{j_3}]}
   {[(n_1/L_1)^2 + (n_2/L_2)^2+ (n_3/L_3)^2]}
|\hat{\psi}(n_1/2^{j_1})\hat{\psi}(n_2/2^{j_2})\hat{\psi}(n_3/2^{j_3})|^2,
\end{eqnarray}
for the last step, we consider that $\hat{\psi}(n_3/2^{j_3})$
requires $n_3\simeq 2^{j_3}n_p$, and $n_p\simeq1$ being the peak
of $\hat{\psi}(n)$. Since $\sum_{l=0}^{2^j-1}\sin(2\pi nl/2^{j})
< 1$, equations.(A7) and (A14) yield
\begin{equation}
 \sum_{l_3-l'_3}Q^2_{{\bf j}, l_3-l'_3} <
  \left (\frac{L_3}{2\pi n_p2^{j_3}}\right )^2 S^2_{\bf j}.
\end{equation}

\newpage

\figcaption {The diagonal DWT power spectrum (square) measured in
simulation samples for models  SCDM  (upper panel), $\tau$CDM
(central panel) and $\Lambda$CDM1 (lower panel). The error bars
are 1-$\sigma$ variance from 10 realizations for each model. The
solid lines are the original power spectrum} \label{Fig1}

\figcaption {The values of $\beta$ estimated by the redshift and
blueshift power spectra in the $\Lambda$CDM2 simulation samples.
The error bars are 1-$\sigma$ variance from 6 realizations. }
\label{Fig2}

\figcaption{$P^S_{jj_2j_3}/P^S_{ jj_3j_2}$ vs. $j$ of
$\Lambda$CDM2 simulation samples. The error bars are 1-$\sigma$
variance from 6 realizations. The solid line is given by a best
fitting with eq.(52)} \label{Fig3}

\figcaption{$S^{PV}_{\bf j}$ vs. $j_{\perp}$ of $\Lambda$CDM2
simulation samples. The error bars are 1-$\sigma$ variance from 6
realizations.} \label{Fig4}

\figcaption{$S^{PV}_{jjj}/S^{PV}_{222}$ of $\Lambda$CDM2
simulation samples. The error bars are 1-$\sigma$ variance from 6
realizations.} \label{Fig5}

\figcaption{PVD estimated from the $\Lambda$CDM2 simulation
samples. The error bars are 1-$\sigma$ variance from 6
realizations for each model.} \label{Fig6}

\figcaption{Reconstructed DWT power spectrum for the $\Lambda$CDM2
simulation samples. The error bars are 1-$\sigma$ variance from 6
realizations.} \label{Fig7}


\begin{thebibliography}{99}

\bibitem{bb} Bardeen, J.M., Bond, J.R., Kaiser, N. \& Szalay,
 A.S., 1986, ApJ, 304, 15

\bibitem{cfw} Cole, S., Fisher, K.B. \& Weinberg, D.H., 1994,
\mnras, 267, 785

\bibitem{c} Couchman, H.M.P., 1991, ApJ, 368, 23

\bibitem{d} Daubechies I. 1992, Ten Lectures on Wavelets,
  (Philadelphia, SIAM)

\bibitem{ft} Fang, L.Z. \& Thews, R. 1998, Wavelet in
 Physics, (World Scientific, Singapore)

\bibitem{ff} Fang, L.Z. \& Feng, L.L. 2000, \apj, 539, 5 (paper I).

\bibitem{fsl} Fisher, K.B., Scharf, C.A. \& Lahav, O., 1994,
 \mnras, 266, 219

\bibitem{H} Hamilton, A.J.S., 1998, The Evolving Universe,
 (Kluwer Academic Publisher, Germany)

\bibitem{hc} Hatton, S.J. \& Cole, S., 1998 \mnras, 296, 10

\bibitem{jbf} Jamkhedkar, P., Bi, H.G. \&  Fang, L.Z. 2001,
   \apj, in press, astr-ph/0107185

\bibitem{k} Kaiser, N. 1987, \mnras, 227, 1

\bibitem{kcpr} Kudlicki, A., Chodorowski, M.J., Plewa, T. \&
  R\'{o}zyczka, M. 2000, \mnras, 316, 464

\bibitem{pd} Peacock, J.A. \& Dodds, S.J., 1994, \mnras, 167, 1020

\bibitem{sml} Szalay, A.S., Matsubara, T. \& Landy, S.D., 1998,
   \apj, 498, L1

\bibitem{yfcf1} Yang, X.H., Feng, L.L., Chu, Y.Q. \& Fang, L.Z.
    2001a, \apj, 553, 1 (paper II).

\bibitem{yfcf2} Yang, X.H., Feng, L.L., Chu, Y.Q. \& Fang, L.Z.
    2001b, \apj, in press (paper III).


\end{thebibliography}
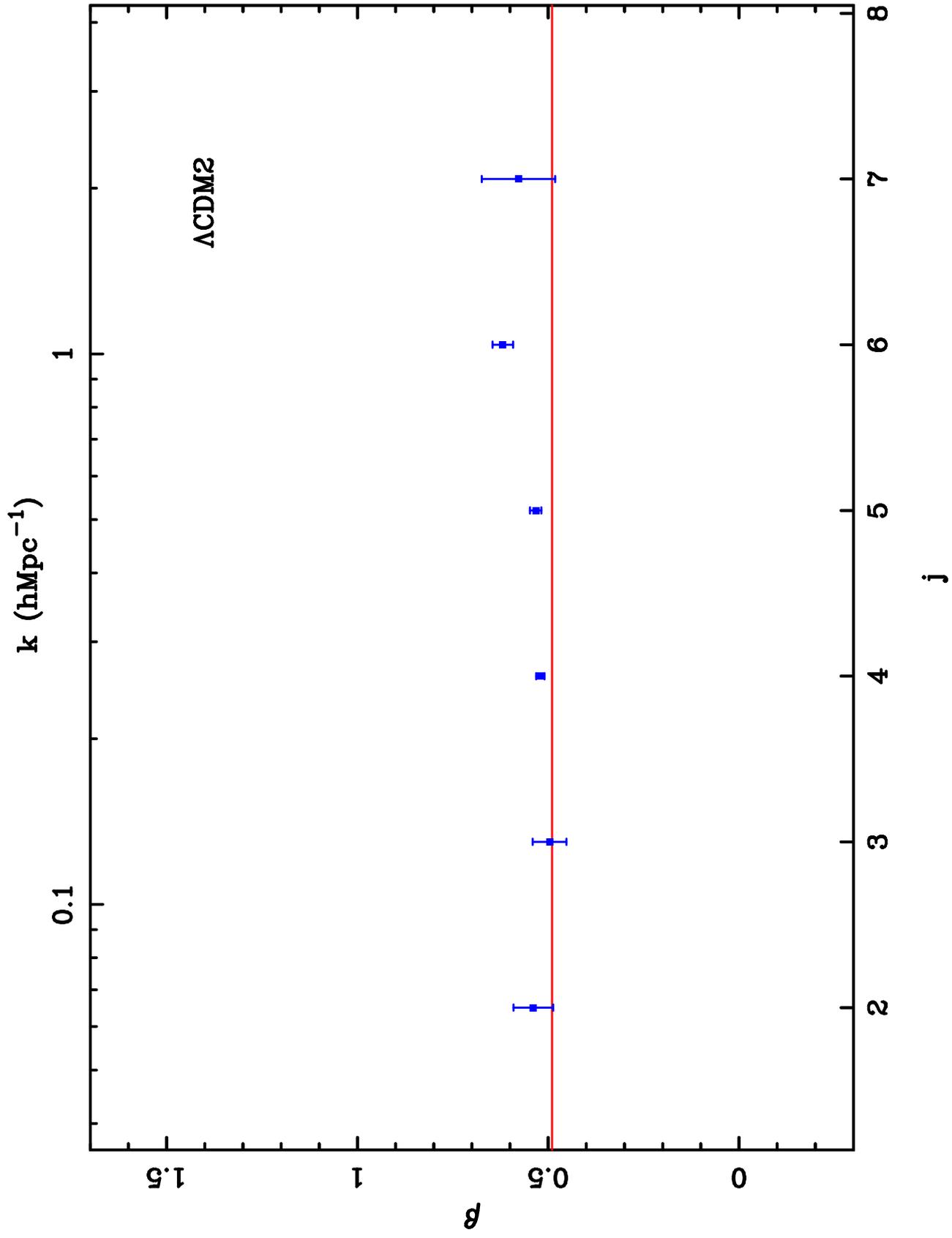
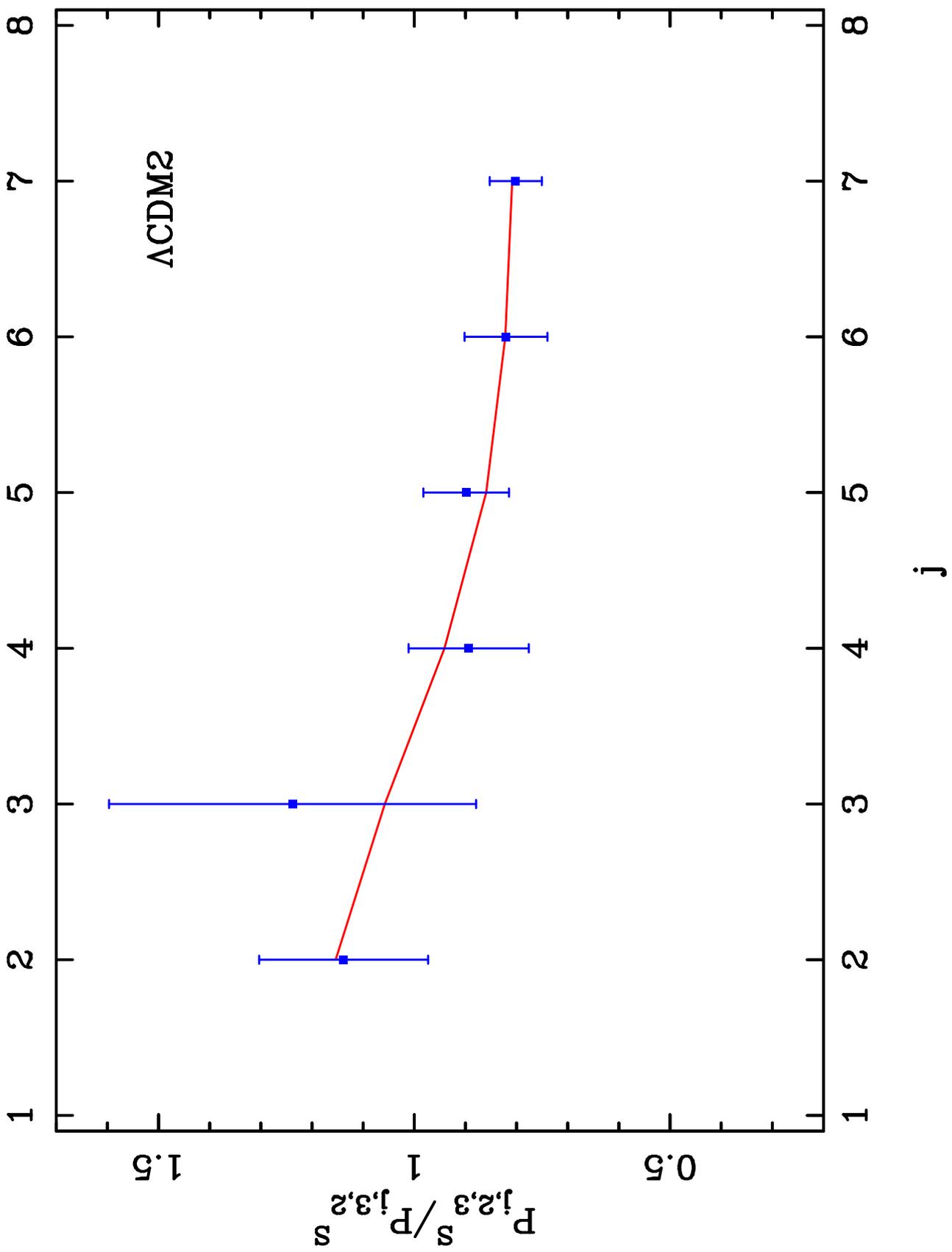
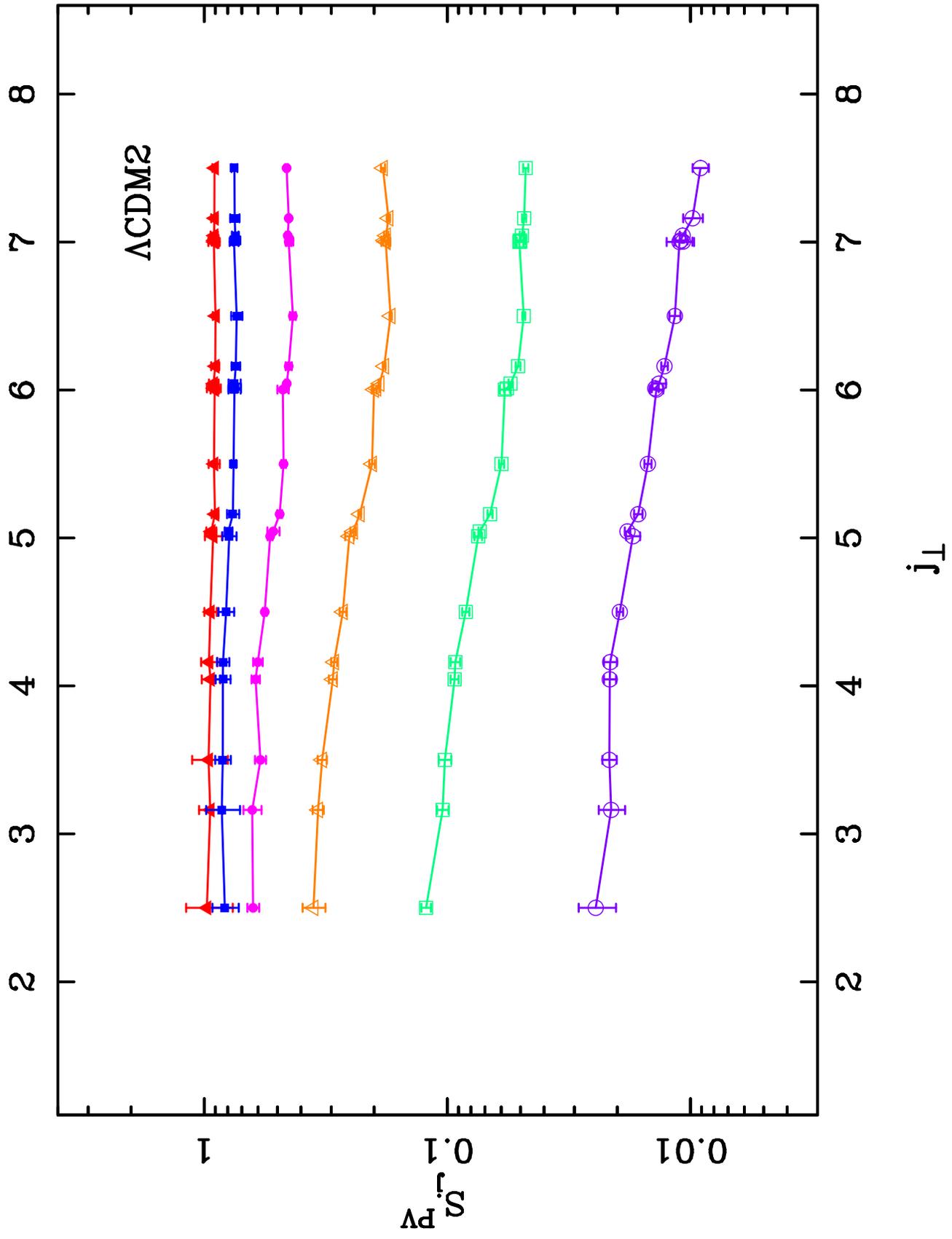
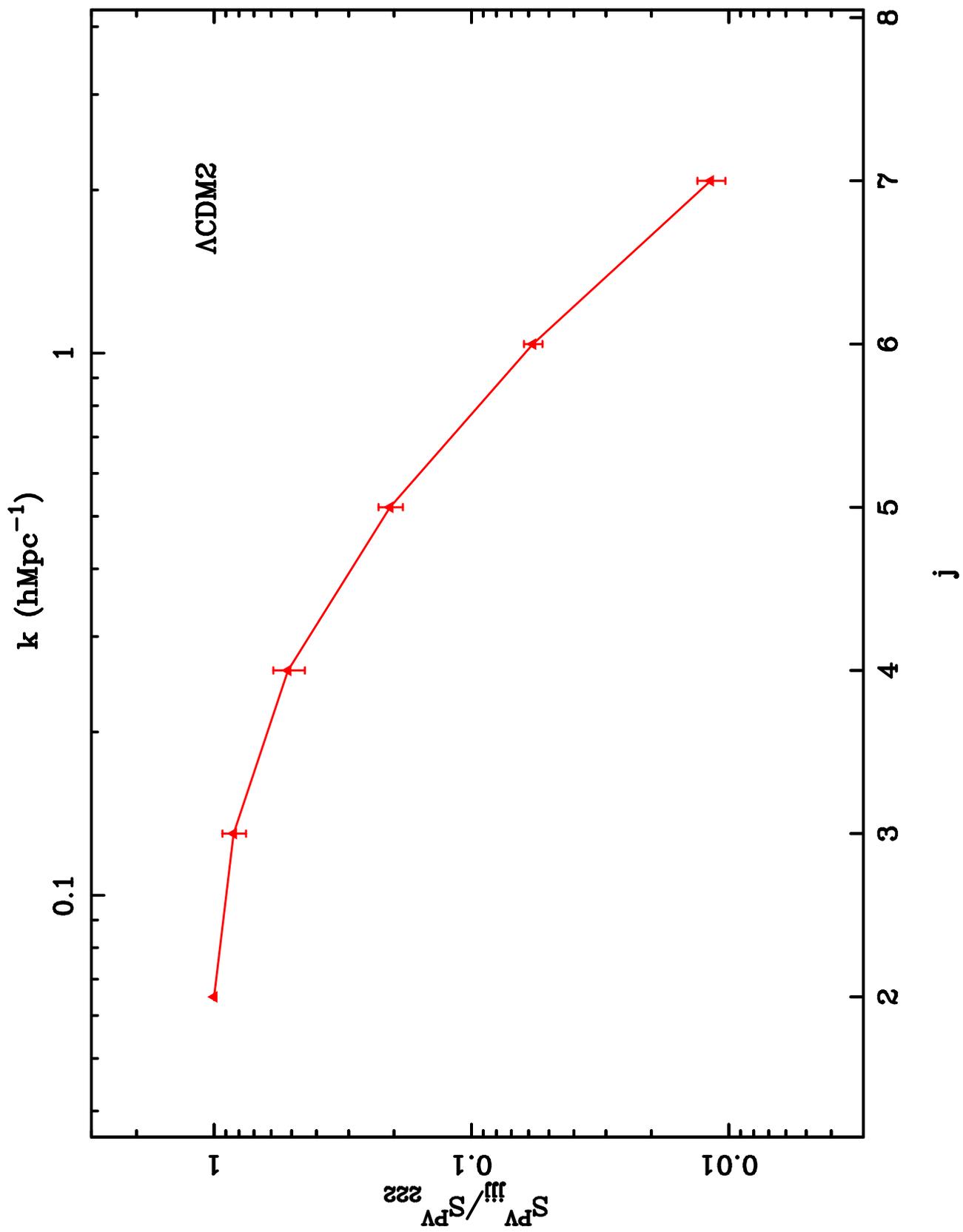
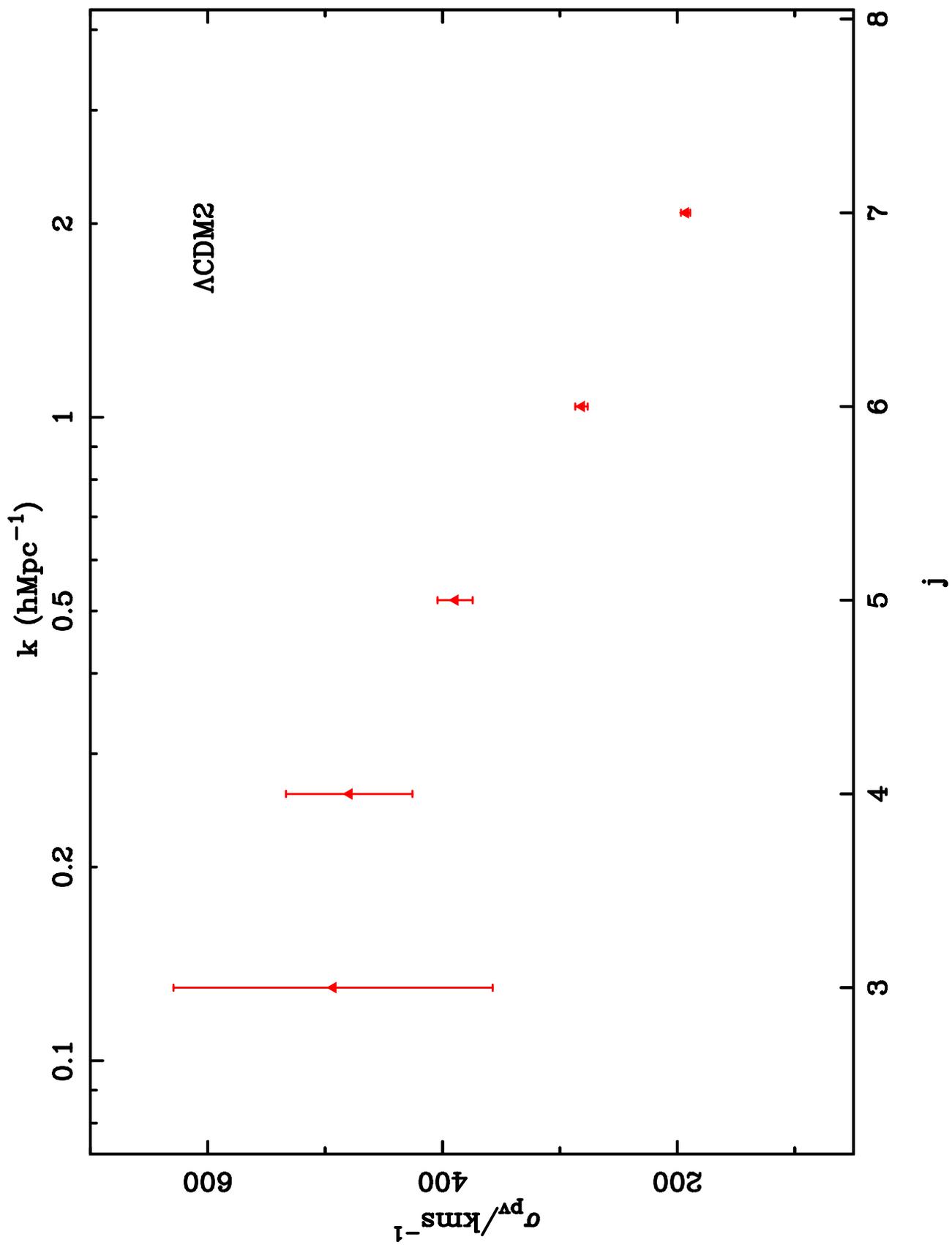
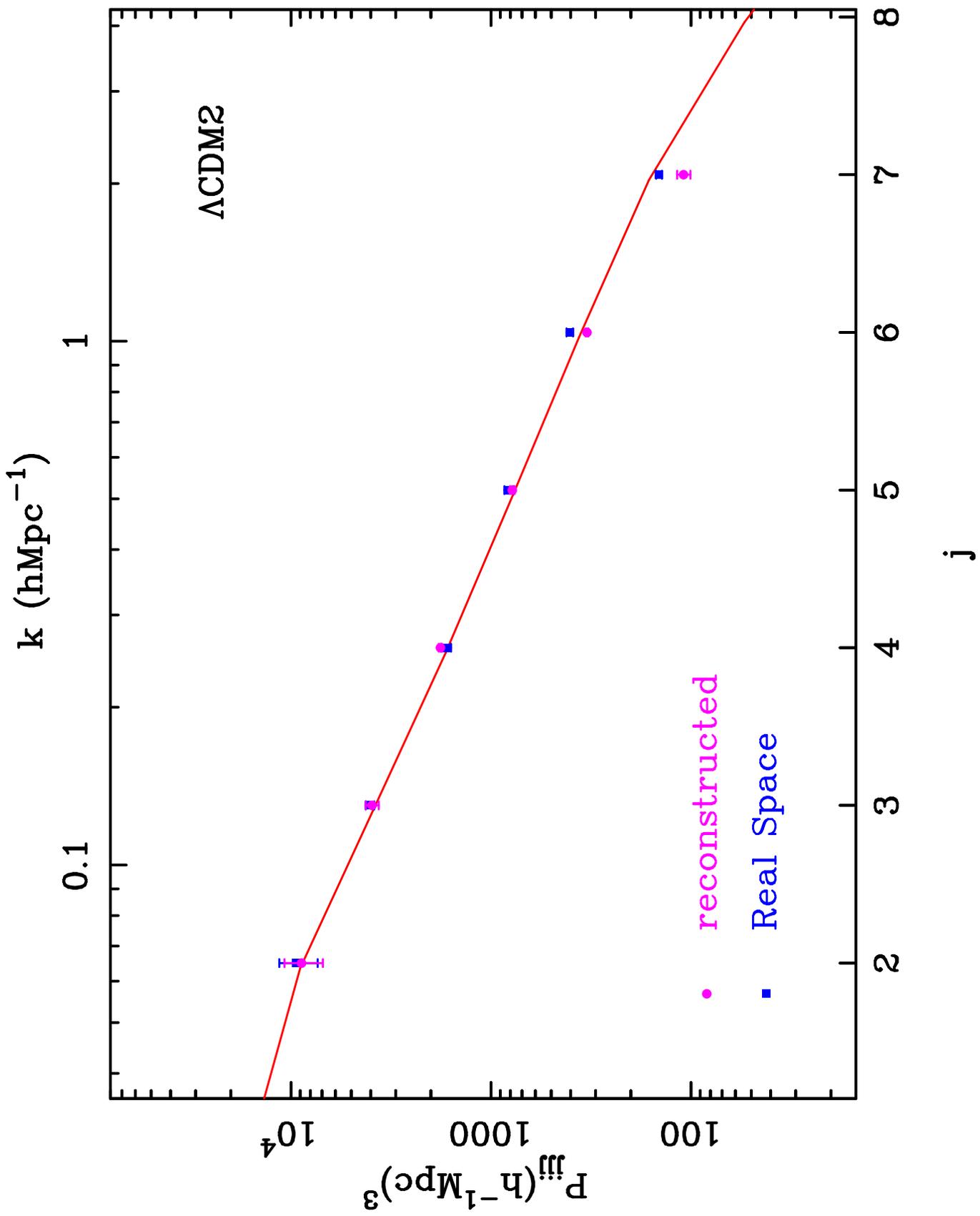
\end{document}